\documentclass[apj]{emulateapj}
\usepackage{apjfonts}
\gdef\msun{M$_{\odot}$}

\begin{document}
\title{The relation between compact, quiescent high redshift galaxies
and massive nearby elliptical galaxies: Evidence for hierarchical,
inside-out growth}

\author{Rachel Bezanson\altaffilmark{1}, Pieter G.\ van Dokkum\altaffilmark{1}, Tomer Tal\altaffilmark{1}, Danilo Marchesini\altaffilmark{1}, Mariska Kriek\altaffilmark{2}, Marijn Franx\altaffilmark{3} and Paolo Coppi\altaffilmark{1}}

\altaffiltext{1}
{Department of Astronomy, Yale University, New Haven, CT 06520-8101}
\altaffiltext{2}
{Department of Astrophysical Sciences, Princeton University, Princeton, NJ
08544}
\altaffiltext{3}
{Sterrewacht Leiden, Leiden University, NL-2300 RA Leiden, Netherlands}

\shortauthors{Bezanson et al.}
\shorttitle{Evidence for Inside-Out Growth of Ellipticals}

\newcommand{\unit}[1]{\ensuremath{\, \mathrm{#1}}}
\slugcomment{Accepted for publication in the Astrophysical Journal}
\begin{abstract}

Recent studies have shown that massive quiescent galaxies at high redshift
are much more compact than present-day galaxies of the same mass.
Here we compare the radial stellar density profiles and the number
density of a sample of massive galaxies at $z\sim 2.3$
to nearby massive elliptical galaxies.
We confirm that the average stellar densities of the $z\sim 2.3$ galaxies
within the effective radius, $\rho(<r_e)$,
are two orders of magnitude higher than those of local elliptical
galaxies of the same stellar mass. However,
we also find that the densities measured
within a constant physical radius of 1 kpc, $\rho(<1\,{\rm kpc})$,
are higher by a factor of 2--3 only. This suggests that inside-out
growth scenarios are plausible, in which
the compact high redshift galaxies make up the
centers of normal nearby ellipticals. The compact
galaxies are common at high redshift, which enables us to
further constrain their
evolution by requiring that the number density of their descendants does
not exceed constraints imposed by the $z=0$ galaxy mass function.
We infer that size growth must be efficient, with $(r_{1+2} / r_1) \sim (M_{1+2} / M_1)^2$.
A simple model where compact galaxies with masses
$\sim 10^{11}$\,\msun\ primarily grow through minor mergers
produces descendants with the approximate sizes,
stellar densities, and number density of elliptical galaxies
with masses $2-3 \times 10^{11}$\,\msun\ in the local
Universe. We note that this model also predicts evolution in the $M_{\rm BH} - \sigma$
relation, such that the progenitors of elliptical galaxies
have lower black hole masses at fixed velocity dispersion.
The main observational uncertainty is 
the conversion from light to mass; measurements of
kinematics are needed to calibrate the masses
and stellar densities of the high redshift galaxies.

\end{abstract}


\section{Introduction}

Several recent studies have found that the oldest and most
massive galaxies at high redshift
have very small sizes
(e.g., {Trujillo} {et~al.} 2006; {Daddi} {et~al.} 2005; {Toft} {et~al.} 2007; {Zirm} {et~al.} 2007; {van Dokkum} {et~al.} 2008; {Cimatti} {et~al.} 2008; {van der Wel} {et~al.} 2008; {Franx} {et~al.} 2008; {Damjanov} {et~al.} 2008; {Buitrago} {et~al.} 2008).
Although these studies use different datasets and methodology
they are in good agreement, finding that the effective radii
of red, apparently quiescent galaxies of fixed mass evolved
by a factor of $\sim 5$ since $z\sim 2.5$
(e.g., {van der Wel} {et~al.} 2008). Initially there were concerns about the
quality of photometric redshifts, the depth of the imaging data,
and the interpretation of the broad-band spectral energy distributions
(SEDs), but these were recently addressed through deep Gemini/GNIRS near-infrared
spectroscopy  and deep HST/NICMOS imaging of a sample of
massive quiescent galaxies at $z\sim 2.3$ ({Kriek} {et~al.} 2006; {van Dokkum} {et~al.} 2008).

These small galaxies are remarkable when compared to nearby galaxies, as
their average stellar densities
are a factor of $\gtrsim 100$ higher than those of red SDSS galaxies of the
same mass ({van Dokkum} {et~al.} 2008). Such massive, dense galaxies are very rare
in the local Universe (e.g., {Trujillo} {et~al.} 2009) but they make up
about half of galaxies with $M\gtrsim 10^{11}$\,\msun\ at $z\sim 2.3$
(e.g, {van Dokkum} {et~al.} 2006; {Kriek} {et~al.} 2006; {Williams} {et~al.} 2008).

Various scenarios have been proposed to explain the observed properties
of the compact galaxies and to describe their subsequent evolution.
The most straightforward explanation is that
the masses are overestimated and/or the sizes underestimated.
The mass measurements currently rely on fitting stellar population
synthesis models to the observed photometry and spectra, and these models
have considerable systematic uncertainties. A significant uncertainty is the
stellar initial mass function (IMF): a ``bottom-light'' IMF, such as
proposed by, e.g., {van Dokkum} (2008), {Dav{\'e}} (2008), and
{Wilkins}, {Trentham}, \&  {Hopkins} (2008), would generally lower the implied
   masses, with the precise effect depending on the age of the stellar populations.
The sizes of the galaxies can be underestimated in several ways. It
may be that the galaxies have strong radial gradients in $M/L$ ratio,
in which case the luminosity-weighted sizes are different from
the mass-weighted sizes (e.g., {Hopkins} {et~al.} 2008).    We also note that {Hopkins} {et~al.} (2008) predict smaller differences
between nearby elliptical galaxies and their
progenitors, due to contribution of the dark
matter halos.  Limitations in resolution and signal-to-noise ratio may also play
a role, although this seems increasingly unlikely.  

Taking the measured masses and sizes at face value, three effects have
been discussed to explain the observed evolution. The first is a
variation on ``progenitor bias'' ({van Dokkum} \& {Franx} 2001),
which states that early-type galaxies at
high redshift are only a subset of all progenitors of today's
early-type galaxies. As we discuss later, the number density of the compact
galaxies at $z\sim 2.3$ 
is only $\sim 7$\,\% of the number density of galaxies
with the same mass today (see \S\,5.1)\footnote{We note that this fraction is smaller than
  that found in Kriek et al.\ (2008). The reason is that we adopt a
   different IMF and therefore a different mass limit, and we
    assume that at $z=0$ all galaxies with $M\ge10^{11}$\,\msun\
  are "red and dead" but only $\sim 50$\,\% at $z=2.5$.}. Therefore, the compact galaxies may
be the progenitors
of the most compact $\sim7$\,\% of today's galaxies with the same mass
(see also {Franx} {et~al.} 2008). This explanation cannot be complete,
as the compact galaxies are small even when
compared to this subset of the present-day population.
The second explanation is minor or major merging, which will increase
the sizes but also the masses ({Khochfar} \& {Silk} 2006; {Naab} {et~al.} 2007; {Hopkins} {et~al.} 2008). Significant
merging is expected for these massive galaxies, e.g., {White} {et~al.} (2007); {Guo} \& {White} (2008),
and merging scenarios have been discussed in several papers
(e.g., {Cimatti} {et~al.} 2008; {van der Wel} {et~al.} 2008). The third explanation that
has been discussed is expansion of the galaxies as a result of
dramatic mass loss due to quasar feedback ({Fan} {et~al.} 2008).

In this paper we provide new constraints on the evolution of compact
``red and dead'' high redshift galaxies. In \S\,3 we compare the radial stellar
density profiles of the compact galaxies to those of nearby elliptical
galaxies. This allows us to determine whether the compact galaxies
resemble the central regions of elliptical galaxies, and hence whether
normal elliptical galaxies are plausible descendants via merging scenarios. In \S\,4 we present three simple models to explain the growth of compact galaxies into local elliptical galaxies.  In \S\,5
we consider which of the modes is most likely to dominate galaxy growth by including constraints from the evolution of the mass function, and derive a lower bound on the amount of
size growth for a given amount of mass growth.  Throughout this paper, we assume a $\Lambda$CDM cosmology with $H_0 = 70 \unit{km} \unit{s^{-1} Mpc^{-1}}$, $\Omega_m= 0.3$,
and $\Omega_{\Lambda} = 0.7$.

\section{Density Profiles}

Density profiles of nearby elliptical galaxies and the
compact high redshift galaxies are constructed.
For the compact galaxies we deproject the Sersic fits
presented in vD08, and for the nearby galaxies we use a
combination of new and literature data.

\subsection{Surface Brightness Profiles}

\subsubsection{High Redshift Galaxies}
We use the sample of nine high redshift ``red and dead''
galaxies previously studied by
{Kriek} {et~al.} (2006) and {van Dokkum} {et~al.} (2008) [hereafter vD08]. The
redshifts of the galaxies were measured from deep rest-frame optical
Gemini/GNIRS spectra ({Kriek} {et~al.} 2006). The spectra also demonstrate
that the light comes from evolved stellar populations,
as they exhibit prominent Balmer or 4000\,\AA\ breaks. The galaxies
were imaged with the Hubble Space Telescope (HST) NICMOS2 camera,
and with Keck/NIRC2 using laser guide star-assisted adaptive optics.
As described in vD08, the galaxies were fit with {Sersic} (1968) profiles
using GALFIT ({Peng} {et~al.} 2002). Structural parameters for the galaxies
are listed in vD08.

Surface brightness profiles in the $H_{160}$ band
were constructed from the Sersic fits. The galaxies are barely resolved
even with the NICMOS2 camera, and we have essentially no information on the
form of the density profile within the effective radius
($0\farcs 1$, or $\approx 1$\,kpc).  The average density within
this radius is much better constrained, and this is the parameter that
we will use in quantitative comparisons. We note that GALFIT effectively
extrapolates the Sersic fits to the (resolved) structure at large
radii inward while conserving the total flux, and that therefore
the fits {\em may} also
provide a good approximation of the form of the density
profile within 1\,kpc.

\subsubsection{Nearby Galaxies}

Two sources are used for the nearby sample.  The {Tal}, {van Dokkum}, \& {Nelan} (2009) [hereafter T09] sample
is an absolute magnitude and volume-limited sample of local elliptical galaxies,
selected from {Tully} (1988). All galaxies with morphological
type ``E'', $M_B < -20$, within
declinations of $-85$ and +10, galactic latitude $> 17^{\circ}$ or $<-17^{\circ}$ and at distances of $15 - 50\unit{Mpc}$
were observed with the Yale 1.0\,m telescope at CTIO,
operated by the SMARTS consortium, in the $V$ band.
The observing strategy was optimized for flat-fielding accuracy,
and the surface brightness profiles can be reliably traced to
$\approx 29$\,mag\,arcsec$^{-2}$.
The galaxies were fit
with isophotal ellipses using IRAF. Apparent magnitudes were
calibrated using aperture photometry of {Prugniel} \& {Heraudeau} (1998) and then
converted to $B$ magnitudes using published $B-V$ colors from the same
catalogue (neglecting color gradients).  Measurements were corrected for Galactic reddening using infrared dust maps from {Schlegel}, {Finkbeiner}, \&  {Davis} (1998).  We assume distance
measurements from the Tully catalogue (corrected to our
cosmology) to convert the luminosity profiles to physical units.

The T09 sample has the advantage that it is complete down to
a luminosity limit (which roughly corresponds to a mass limit for
these luminous red ellipticals), but the disadvantage is that it
only samples a limited range in mass and luminosity. We supplemented the
T09 sample with photometry from {Franx}, {Illingworth}, \&  {Heckman} (1989); {Peletier} {et~al.} (1990); {Jedrzejewski} (1987) [hereafter FPJ]. This
sample is not complete but covers a larger range in luminosity.
We limited the sample to all galaxies that have
published $B$-band profiles.  Again distances from {Tully} (1988) were used to convert the
observed brightnesses to luminosities.

\subsection{Deprojection}
The intensity profiles of the nearby galaxies
are fit to Sersic profiles of the form
\begin{equation}
  I(r) = I_o \exp{\left[-b_n {\left(\frac{r}{r_e}\right)}^{1/n}\right]}
\end{equation}
with $n \le 4$ between radii of $4\unit{''}$ out to $20\unit{kpc}$, or
the maximum extent of each profile, along the circularized axis, $r =
a\sqrt{(1-\epsilon)}$, of the galaxy ({Ciotti} 1991).  $b_n$ is defined
as the solution to $\gamma(2n,b_n) = \Gamma(2n)/2$. We
use the asymptotic approximation for $b_n$, which is accurate to a
factor of $\mathcal{O} \sim 10^{-6}$:
\begin{equation}
b_n \approx 2n - \frac{1}{3}+\frac{4}{405n}+\frac{46}{25515n^2}
\end{equation}
({Ciotti} \& {Bertin} 1999). For the high redshift galaxies we used the
fits of vD08. 
We then performed
an Abel Transform to deproject a circularized, three-dimensional light profile:
\begin{equation}
  \rho_{L}\left(\frac{r}{r_e}\right) =
  \frac{b_n}{\pi}\frac{I_o}{r_e}{\left(\frac{r}{r_e}\right)}^{(1/n-1)}
  \int_{1}^{\infty}\frac{\exp{[-b_n{(\frac{r}{r_e})}^{(1/n)}t]}}
  {\sqrt{t^{2n}-1}}dt
  \end{equation}
For both the high redshift sample of compact galaxies and the two
samples of nearby elliptical galaxies we now have 
circularized radial luminosity density profiles in units of
$L_{B,\odot}\,{\rm kpc}^{-3}$.

\subsection{Light-to-Mass Conversions}
In order to convert the luminosity density profiles to stellar mass density
profiles we make the following assumptions
about the mass-to-light ($M/L$) ratios. For the high redshift sample
we use stellar masses from {Kriek} {et~al.} (2008) adjusted to a
{Kroupa} (2001) IMF. For the nearby sample, we use the well-established
relation between $M/L$ ratio and mass to convert luminosities to
masses (e.g., {van der Marel} 1991). The normalization and slope
of the relation in the $B$ band were determined by combining the
information in Table 1 of {van der Marel} (1991) and Table 2
in {van der Marel} \& {van Dokkum} (2007):

\begin{equation}
\frac{M}{L_B} = (9.04\times 10^{-4}) {\left(\frac{L_{B}}{L_{B,\odot}}\right)}^{0.37}
\end{equation}

The conversion from luminosity to mass is the largest uncertainty in
the methodology, in particular the lack of dynamical measurements
that could calibrate the $M/L$ ratios of the high redshift galaxies.
We will return to this issue in \S\,6.

\section{Comparison of Density Profiles}

\subsection{Average Profiles}
\begin{figure}[h]
  \centering
  \includegraphics[scale=0.85]{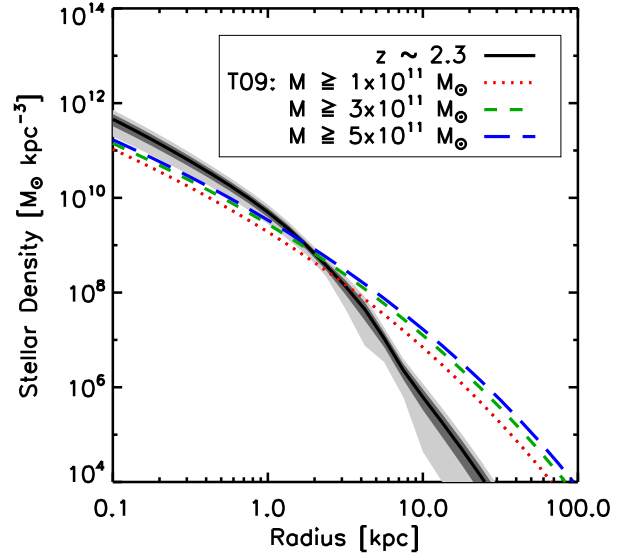}
  \caption{Comparison
 of the mean stellar density profiles of high redshift compact
galaxies
    (solid line) with nearby elliptical
galaxies from the T09 sample (broken lines).  High redshift 1 and 2 $\sigma$ contours are shown in gray.  The average
    density profile for nearby galaxies with $M \ge 10^{11}M_{\odot}$ is
    represented by the red, dashed line; the green, short dashed line
corresponds to galaxies with $M \ge 3\times
    10^{11}M_{\odot}$; and the
    most massive local galaxies with $M \ge 5 \times 10^{11}
    M_{\odot}$ are shown by the blue, long-dashed line. Note
that the profiles of the nearby galaxies are fairly similar to
those of the compact galaxies at radii $r\lesssim 3$\,kpc, qualitatively
consistent with expectations for inside-out growth.}
  \label{fig:densprof}
\end{figure}

The stellar density profiles of the compact high redshift galaxies are compared
to those of nearby elliptical galaxies in Fig.\ \ref{fig:densprof}.
The solid line is the average density profile of the 9 galaxies from vD08. We use a 1000 iteration bootstrap estimation to approximate errors of the average density profile due to the small sample size of the high redshift galaxies.  The 1 $\sigma$ contour is shown in dark gray and the 2 $\sigma$ is shown in light gray.  Broken lines are average
profiles of nearby elliptical galaxies from the T09 sample, in
three different mass bins. The lowest mass bin is
$M \ge 10^{11}M_{\odot}$: this is the mass that the high redshift
galaxies already have at the epoch of observation, and therefore
the minimum mass of their descendants.

Figure \ref{fig:densprof} shows that the discrepancy between
the profiles of compact high redshift galaxies and nearby elliptical
galaxies is mostly in the outer regions.
Within $r \approx 1\unit{kpc}$
the average stellar density of the high redshift galaxies is
greater than the density of nearby ellipticals by a
factor of a few only, particularly for
the more massive galaxies in the T09 sample.  This
discrepancy is much smaller than the factor of $\gtrsim 100$
difference when the density is measured within the effective
radius (e.g., vD08).  Furthermore, our error estimates only address the sample bias; this discrepancy is especially small considering the other sources of uncertainty in our measurements, which we will discuss further in \S\,6.
Outside of this inner region, the difference
grows significantly; the stellar density of nearby elliptical
galaxies is a few hundred
times higher than that of the compact high redshift galaxies
at $r>10 \unit{kpc}$. We infer that in order to
evolve into nearby galaxies, compact galaxies need not change
significantly in the central regions, but must grow significantly in
their outer regions.  

\begin{figure*}[t]
  \centering
  \includegraphics[scale=0.9]{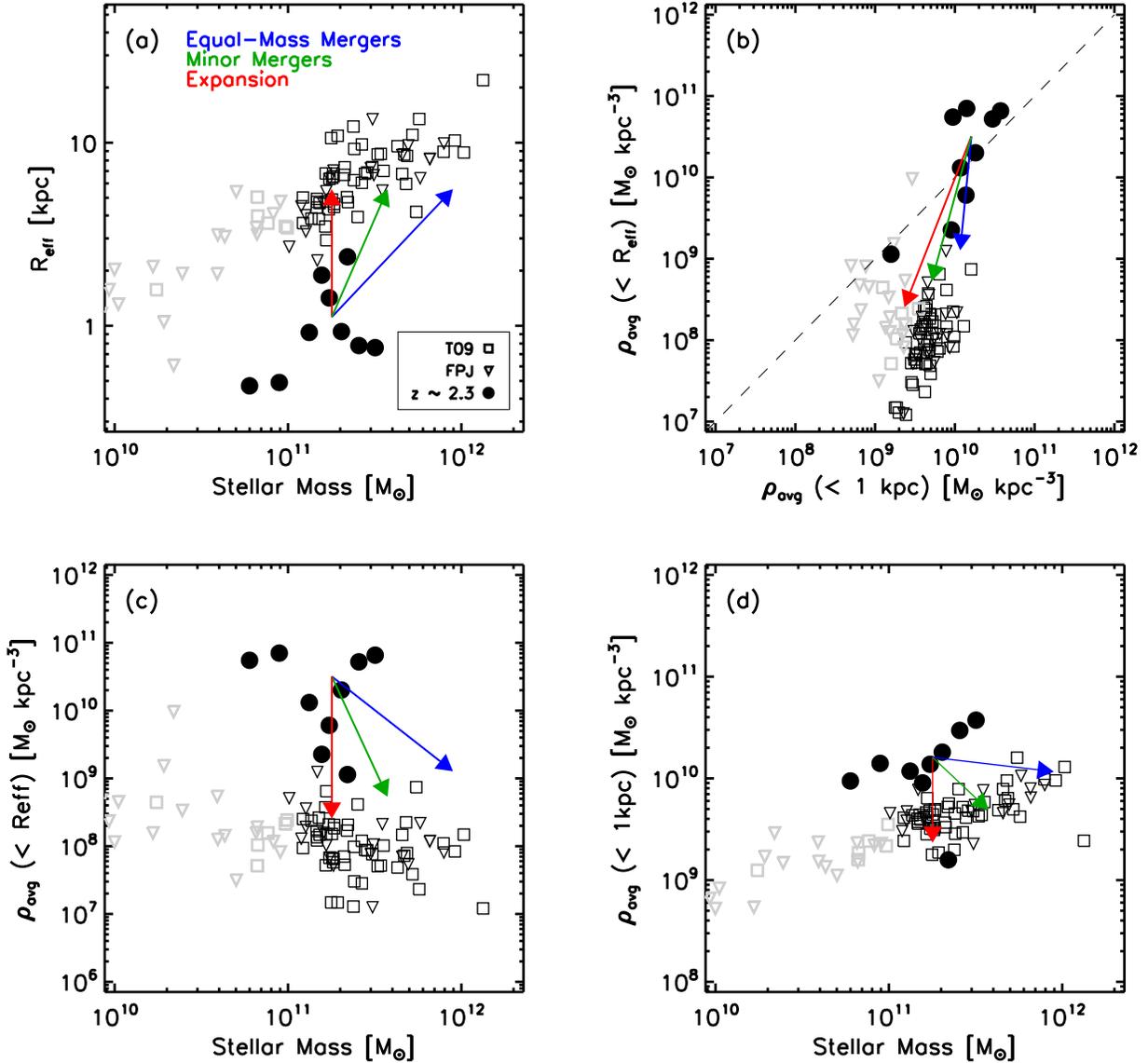}
  \caption{Relative properties of nearby and $z\sim2.3$ galaxies.
The panels show the relations between size and mass (a),
density within the effective radius and density within 1\,kpc (b),
density within the effective radius and mass (c), and density
within 1\,kpc and mass (d).
Open symbols are nearby galaxies, solid circles are high redshift compact
galaxies from vD08. Light grey points are 
nearby galaxies with masses $< 10^{11}\unit{M_{\odot}}$, i.e.,
lower than the high $z$ compact galaxies. Arrows begin at mean values of high redshift sample and show predictions
from simple models for the evolution of the compact galaxies: blue arrows shows the direction of evolution due to equal-mass mergers, green arrows for minor mergers and red arrows for the expansion model.
Simple expansion or minor mergers can bring the distant
galaxies close to the scaling relations defined by nearby galaxies,
but equal-mass mergers do not produce galaxies of the right size.}
  \label{fig:allprop}
\end{figure*}

\subsection{Comparison of Masses, Sizes, and Densities}

The relative properties of the high redshift galaxies and nearby
galaxies are demonstrated in Figure \ref{fig:allprop}. The compact
galaxies from vD08 are indicated by solid circles. The nearby
samples are represented by open symbols: squares for T09 and triangles
for FPJ.  Only nearby galaxies with sufficient mass,
$M \gtrsim 10^{11} M_{\odot}$, can be the descendants of the high redshift
galaxies. Galaxies
with lower masses are denoted with light grey symbols.

The relative compactness of high redshift and low redshift galaxies is shown in Figure
\ref{fig:allprop}(a). There is a clear trend showing the increasing
effective radius with galaxy mass in the nearby galaxies. The high
redshift galaxies, though in the middle of the nearby mass range, are
smaller by a factor of $\sim 5$ in effective radius. This result
confirms previous studies, which generally used the Sloan Digital
Sky Survey (SDSS) as a low redshift comparison point
(e.g., {Toft} {et~al.} 2007; {van Dokkum} {et~al.} 2008; {Cimatti} {et~al.} 2008; {van der Wel} {et~al.} 2008). {van der Wel} {et~al.} (2008)
combine data from the literature (in addition to adding
new data at $z\approx 1$) and derive an evolution of $r_e
\propto (1+z)^{-1.20\pm 0.12}$ at fixed mass
(for samples with photometrically determined masses), corresponding to
a factor of $4.2\pm 0.6$ at $z=2.3$.

The difference in size at fixed mass implies a
significant difference in density contained within the effective
radius of the high redshift and nearby galaxies. We calculated the
average densities within the effective radius by integrating the
stellar density profiles derived in the previous Section:
\begin{equation}
\rho(<r) = \frac{3}{r^3} \int_0^{r} \rho(r'){r'}^2 dr'
\end{equation}
with $r=r_e$.
This difference is
obvious in Figure \ref{fig:allprop}(b): the vertical
axis of this panel demonstrates the factor of $>100$
differences in the average density within the effective radius.

The horizontal axis of Fig.\ \ref{fig:allprop}(b) shows
the average density integrated to $r=1$\,kpc rather than
$r=r_e$. For convenience, we will refer to the average density
within 1\,kpc as the ``central density''.
Since the compact galaxies have effective radii $\sim 1 \unit{kpc}$, the
density within $r_e$ is approximately equivalent to our definition
of the central density, placing these galaxies along the dashed
diagonal line representing the equality of the two densities. The
nearby sample lies predominantly below this line, with $\rho(<
r_e)$ much lower than $\rho(<1\,{\rm kpc})$ for all galaxies with
masses $>10^{11}$\,\msun.
We infer that, although the high redshift galaxies have higher
densities than nearby ellipticals overall, the differences
are much smaller within 1 kpc than within $1\,r_e$.

Figure \ref{fig:allprop}(c) and (d) demonstrate the same
point in the density versus mass plane.
In (c), we show the relation between $\rho(<r_e)$ and
total stellar mass. The compact high redshift galaxies are
clearly much denser than nearby galaxies of the same mass.
In (d), it is shown that the 
discrepancy in density becomes far less extreme
in the central regions of the galaxies.
The nearby sample shows opposite trends with mass in (c) and (d):
the density within the effective radius decreases with increasing
mass (reflecting the slope of the mass -- radius relation), but
the density within 1\,kpc grows with increasing mass.
Interestingly, the central densities of the high redshift compact
galaxies are very similar to those of nearby elliptical galaxies
with masses $\gtrsim 5 \times 10^{11}$\,\msun.

The trends in Fig.\ \ref{fig:allprop} are
consistent with models in
which the compact galaxies make up the centers of
present-day giant ellipticals. Such inside-out formation scenarios are not
new, and have been explored by, e.g., {Loeb} \& {Peebles} (2003), {Bournaud}, {Jog}, \&  {Combes} (2007), {Naab} {et~al.} (2007),
and {Hopkins} {et~al.} (2008).  The idea is that a compact core is formed through highly dissipative
processes at $z\gtrsim 3$ (see, e.g., {Robertson} {et~al.} 2006a; {Dekel} {et~al.} 2008),
which then grows through increasingly
dissipationless mergers at lower redshift.  Independently, Franx et al. 2008 argues that galaxy growth is mostly
inside-out, based both on the regular evolution of the stellar mass-radius
relation, and on the fact that star forming galaxies are larger than
non-star forming galaxies of the same mass.

\section{Predictions from Simple Models}

As discussed in \S\,1, various models have been proposed to
explain the apparent growth of massive galaxies since $z\sim 2.5$.
Here we discuss three possible simple models in the context of
the relations shown in Fig.\ \ref{fig:allprop}:
equal-mass mergers, minor mergers and expansion at fixed
mass. We investigate the effects of these models in Fig.\ \ref{fig:allprop}
with arrows. The starting point of the arrows is always the mean of the
high redshift compact galaxies, and they all imply a growth
in effective radius of a factor of 5.  We emphasize that we look to constrain the dominant mode of galaxy evolution; while individual galaxies in the sample will likely be affected by all of the processes discussed below, we focus on the overall trends in the larger context of the sample of galaxies.

\subsection{Model 1: Growth via Equal-Mass Mergers}

In this model, the growth is driven by
(near-) equal mass mergers. These mergers will not only increase
the size of the galaxies, but also their mass. Applying straightforward
virial arguments implies
\begin{equation}
K_{1+2} = K_1 + K_2,
\end{equation}
with $K_{1+2}$ the kinetic energy of the remnant and $K_1$,
$K_2$ the kinetic energy of the progenitors. With $K= \frac{1}{2}M\sigma^2$
we have
\begin{equation}
\frac{1}{2} M_{1+2} \sigma_{1+2}^2 = \frac{1}{2}M_1\sigma_1^2 +
 \frac{1}{2}M_2 \sigma_2^2,
\label{eq:virial}
\end{equation}
and as $M_{1+2} = M_1+M_2$ and $M_1=M_2$,
we have $\sigma_{1+2}^2 = \sigma_1^2$.
Using $\sigma^2 \propto GM/r$, we arrive at
\begin{equation}
\frac{r_{1+2}}{r_1} = \frac{M_{1+2}}{M_1},
\end{equation}
the familiar result that mergers lead to an increase in size and
mass but no change in velocity dispersion (e.g., {Barnes} 1992). 
We note that these relations are simplifications,
which are inconsistent with the observed slopes of the
stellar mass -- radius relation and the stellar mass -- $\sigma$
relation. Simulations
which take the initial orbits and effects of energy transfer
to the dark matter halos into account generally imply
a smaller increase in size for a given change in mass.  
{Boylan-Kolchin}, {Ma}, \&  {Quataert} (2006) find that $r_{1+2}/r_1 \sim (M_{1+2}/M_1)^{0.6-1}$,
depending on the orbital configuration.

The blue arrows in Fig.\ \ref{fig:allprop}(a-d) show the effects
of equal-mass mergers on the various relations between mass, size,
and density. The density within 1\,kpc was calculated by assuming
that the Sersic indices of the profiles of the compact galaxies do not change.
The blue arrows imply that the descendants of the compact galaxies
are the dominant galaxies in massive groups and clusters,
with stellar masses of $\sim 10^{12}$\,\msun.  As can be
seen in panel {\em d}, the central
densities of these galaxies are nearly identical to those
of the compact galaxies. However, as can be seen
in panel {\em a}, the effective radii of these
giant, nearby galaxies are
a factor of $\sim 10$ larger than the compact objects,
not a factor of $\sim 5$. Therefore,
this model is not  a very good description of the
required evolution in panels {\em a} -- {\em c}.

\subsection{Model 2: Growth via Minor Mergers}

In this mode of galaxy growth, the progenitor galaxies
accumulate mass via minor mergers with small systems. 
The difference with the equal-mass merger model is that minor mergers
are more effective in ``puffing up'' the size a galaxy for a given change
in stellar mass.  For minor mergers $\sigma_1^2 \gg \sigma_2^2$ in
Eq.\ \ref{eq:virial}, and therefore
\begin{equation}
\frac{\sigma_{1+2}^2}{\sigma_1^2} \approx \frac{M_1}{M_{1+2}}
\end{equation}
Again using $\sigma^2 \propto GM/r$ we have
\begin{equation}
\frac{r_{1+2}}{r_1} = \left(\frac{M_{1+2}}{M_1}\right)
 \left( \frac{\sigma_1^2}{\sigma_{1+2}^2}\right)
 \approx \left(\frac{M_{1+2}}{M_1}\right)^2.
\end{equation}
The effective radius grows by the square of the change
in mass (rather than linearly, which is the case for
equal-mass mergers) and the velocity dispersion decreases by the
square root of the change in mass (rather than
remaining constant) (see also {Naab} {et~al.} 2009). As an example, eight successive
$M_2:M_1=1:10$ mergers could lead to a factor of $\sim 5$
increase in effective radius, while the mass would grow
by a factor of $\sim 2$ only.

The effects of this scenario are shown by the green arrows in Fig.\
\ref{fig:allprop}. Again, the density within 1\,kpc was
calculated by assuming that the Sersic index of the profiles
remains unchanged. The compact galaxies have a median mass of $1.7\times10^{11}$\,\msun\, therefore the minor merger model predicts that their
descendants are in galaxies with
a median mass of $3-4\times 10^{11}$\,\msun\ today. The central
densities of
these galaxies are a very good match to those of the
predicted descendants (panel {\em d}), and the effective radii are a much
better match than in the equal-mass merger model (panel {\em a}).  
We note here that what matters is the direction of the arrows, as
their length is arbitrarily
determined by a growth of a factor of five in $r_e$. Extending
the green arrows slightly would bring them very close to the distribution
of nearby elliptical galaxies in all panels.

\subsection{Model 3:  Expansion at Fixed Mass}

In the final model that we examine, a galaxy has accumulated
most of its mass by $z \sim 2$ and then gradually expands over time
while its mass stays roughly constant.
The motivation for this class of models was provided by
{Fan} {et~al.} (2008); they suggest that a QSO may blow out a large fraction of the mass, leading to a significant "puffing up" of the remnant.  We will discuss whether such models are physically
plausible in \S\,5.2 (see also {Trujillo} {et~al.} 2009).

Predictions for these models are indicated by red arrows in Fig.\
\ref{fig:allprop}. By construction, these models predict the right
amount of size evolution at fixed mass, and therefore produce
the same values of $\rho(<r_e)$ as nearby elliptical galaxies.
However, as can be seen in panels {\em b} and {\em d}
they slightly under-predict the central densities of local
ellipticals. The values of $\rho(<1\,{\rm kpc})$ that
are predicted are a factor of $\sim 2$ lower than those of
local ellipticals with the same mass. 

\subsection{Summary of Model Comparisons}

We conclude that all three simple models bring the compact galaxies
much closer to the relations defined by nearby elliptical galaxies.
The equal-mass merger model provides the worst description of the three
as it does not produce elliptical galaxies of the right size and is therefore probably not the dominant mode of growth. The
minor merger is more effective in puffing up the compact galaxies
and, despite its simplicity, provides a remarkably good description of
the masses and densities of nearby elliptical galaxies. The expansion
model provides a good description as well, although it slightly
under-predicts the central densities of elliptical galaxies.

\section{Discussion}

The main result of the previous Section, and this paper, is that the
properties of high redshift compact galaxies can be reconciled with
those of nearby massive elliptical galaxies. The densities
of the compact galaxies are similar to the central densities
of elliptical galaxies, and simple ``toy models'' can be used to
describe the evolution.

\subsection{Independent Constraints on Mass Growth}

The amount of mass
growth in the models of \S\,4 is specified by the physical mechanism
for growth (equal-mass merger, minor mergers, and expansion) and the
choice of a factor of five increase in effective radius. To achieve this
increase, the equal-mass merger model requires an increase in mass of
a factor of $\sim 5$, the minor merger model requires an increase of
a factor of 2--3, and the expansion model does not require an
increase at all.

The evolution of the galaxy mass function provides an independent constraint
on the mass growth. The compact galaxies are common at the epoch of
observation -- they constitute
$\gtrsim 90$\,\% of ``red and dead'' galaxies at $z=2.3$
(vD08) and therefore some $\sim 50$\,\% of the general population of
galaxies with stellar masses $\gtrsim 10^{11}$\,\msun\
(e.g., {Kriek} {et~al.} 2006; {Kriek} {et~al.} 2008; {Williams} {et~al.} 2008). The evolution
of the galaxy mass function
has been measured recently by several groups (e.g., {Drory} {et~al.} 2005; {Fontana} {et~al.} 2006; {Marchesini} {et~al.} 2008; {P{\'e}rez-Gonz{\'a}lez} {et~al.} 2008) . Here, we use
the results from {Marchesini} {et~al.} (2008), who have combined data from
both deep and wide surveys in a self-consistent way.
Integrating the Schechter
function fit given in {Marchesini} {et~al.} (2008) we find that
the integrated number density of galaxies with stellar masses
$M>10^{11} \unit{M_{\odot}}$ is $7.2^{+1.1}_{-1.1}
 \times 10^{-5}\unit{Mpc^{-3}}$ at $z=2.5$. The number
density of compact,
quiescent
galaxies is therefore $3.6^{+0.9}_{-0.9}\times 10^{-5}\unit{Mpc^{-3}}$,
where we assumed that the quiescent fraction is $0.5 \pm 0.1$.
The stellar mass density in these galaxies is $4.8^{+1.6}_{-1.9}
\times 10^6$\,\msun\,Mpc$^{-3}$.

\begin{figure}[t]
\centering
\includegraphics[scale=0.9]{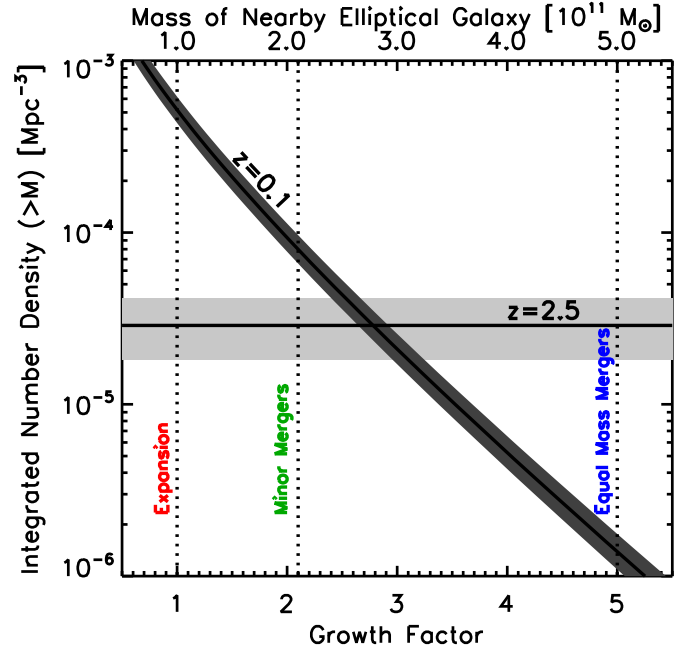}
\caption{Integrated number density of galaxies above a mass limit. The horizontal
line is the number density of quiescent galaxies with $M>10^{11}$\,\msun\
at $z=2.5$. The diagonal relation is the number density at low redshift
as a function of the mass limit. The mass limit is indicated in absolute
units on the top axis, and as a growth factor compared to $z=2.5$
on the bottom axis. Vertical lines
indicate the growth implied by the simple models discussed in \S\,4.
Ignoring merging of compact galaxies with themselves, the mass functions at $z=2.5$ and $z\sim0$ do not allow for growth of more than a factor of 2--3.  Strong merging of compact galaxies is ruled out by the integrated mass density at low redshift (see text).
}
\label{fig:numdens}
\end{figure}

This number density and mass density provide strict bounds on the
typical masses of the descendants of the compact galaxies. If the
mass density of the compact galaxies exceeds that of
local galaxies of a particular mass
it is immediately clear that these local galaxies 
cannot constitute the (sole) descendants.  Figure
\ref{fig:numdens} shows the integrated Schechter stellar mass function
in the local universe in dark grey, as well as the number density of
compact galaxies with $M>10^{11}M_{\odot}$ at $z=2.5$ in light grey.
For any
descendant population at $z=0.1$, mass corresponds to a required growth factor,
given on the lower axis.  We first ignore mergers of compact galaxies with themselves and address that possibility later.  In order for a model to be a feasible
evolutionary path the implied descendant population of galaxies in
the local universe must be at least as common as the progenitors at
high redshift.

The number density of nearby massive galaxies limits the mass
growth to a factor of $2- 3$. For this mass growth each compact
galaxy has one descendant. Lower mass growth implies that
only a small fraction of massive galaxies today hosts a descendant
of a compact galaxy. Higher mass growth is not allowed, as it
would create too many descendants.  Of course, these constraints are dependent on the masses at $z=2.5$, which we derived.  If these masses are incorrect, this argument might change.

Vertical lines in Fig.\ \ref{fig:numdens} indicate predictions
from the three models discussed in \S\,4. The expansion model
is obviously fully consistent with the constraints imposed
by the evolution of the mass
function, as it implies no mass growth. We note that only $\sim 7$\,\%
of nearby galaxies with
masses $>10^{11}$\,\msun\ are descendants of quiescent $z=2.5$
galaxies in this model; we will return to this point below.
Remarkably, we can rule out the equal-mass merger model as the main mode of growth based on
Fig.\ \ref{fig:numdens}, as it implies a mass growth of a factor
of $\sim 5$. The number density of nearby galaxies
with  $M>5\times 10^{11}$\,\msun\ is lower by more than
an order of magnitude than the number density of compact galaxies
with $M>10^{11}$\,\msun\ at $z=2.5$.  In the equal-mass merger model, compact galaxies can obviously merge with each other, which will lower their number density.  However, a factor of $\sim 5$ mass growth is not allowed even when compact galaxies are {\it only} permitted to merge with each other:
the stellar mass density in galaxies with $M>5\times 10^{11}$\,\msun\ at
$z=0.1$ is $8.1^{+2.1}_{-1.6}\times 10^5$\,\msun\,Mpc$^{-3}$,
a factor of 6 lower than the mass density in compact galaxies
with $M>10^{11}$\,\msun\ at $z=2.5$.

Also remarkably, the growth in the minor merger model is close to the cross-over
point, where each compact galaxy has one descendant. 
A plausible explanation is that the central parts of many
elliptical galaxies formed at $z>2.5$, after which they
grew through minor, mostly dry mergers.

More generally, we can combine panel {\em a} of Fig.\ \ref{fig:allprop}
with Fig.\ \ref{fig:numdens} to derive an empirical constraint on the
amount of size growth for a given amount of mass growth. Parameterizing
the relation between size growth and mass growth as
\begin{equation}
\frac{r_{1+2}}{r_1} = \left(\frac{M_{1+2}}{M_1}\right)^{\alpha},
\label{eq:sizemass}
\end{equation}
we find that $\alpha \gtrsim 2$ to simultaneously satisfy the constraints
from the evolution of the size -- mass relation (Fig.\
\ref{fig:allprop}{\em a}; van der Wel et al.\ 2008),
and from the evolution of the mass function. This limit for $\alpha$
is similar to naive expectations from minor mergers, which is
why we obtain a good correspondence between progenitors
and descendants for this class of models. The equal-mass merger
model has $\alpha \sim 1$ (or even $<1$; see {Boylan-Kolchin} {et~al.} 2006);
for the expansion model $M_{1,f}/M_{1} = 1$ (or even $<1$ ) and Eq.\ \ref{eq:sizemass} is
not well defined.

\subsection{Which Models are Physically Plausible?}

We expect that each of these toy models is responsible at some level for the growth and evolution of galaxies from $z \sim 2.5$ until today.  Observational evidence of merging events, both equal-mass and minor, exists at intermediate redshifts and can definitely produce growth in galaxy mass and size.  Mass loss from the central regions of galaxies should also occur and would therefore cause increases in galaxy sizes.  Given this complexity, we hope to identify which of our simple models best describes the mechanism responsible for the majority of the growth of the compact galaxies at high redshifts into descendant galaxies in the nearby universe.

In \S\,5.1 we found
that the equal-mass merger model is inconsistent with the number density
of massive galaxies today.
We are therefore left with two feasible models, growth via ``in-situ''
expansion or via minor
mergers.  Both of these modes of galaxy growth have the effect of
puffing up the galaxies without extreme mass growth.  Number densities
of the implied descendants of galaxies that have grown via either mode
correspond to sufficiently common galaxies in the local universe.

Although number density arguments do not immediately
discredit the expansion
model of galaxy evolution, they do lead to uncomfortable questions.
The implication of no mass growth is that only a very small number
of nearby galaxies with mass $>10^{11}$\,\msun\ was already formed
at $z=2.5$: approximately 7\,\% if only quiescent galaxies at $z=2.5$
are considered, and $\sim 14$\,\% if all galaxies with
$M>10^{11}$\,\msun\ are considered. This raises the question
where the progenitors of the remaining $\sim 90$\,\% of today's massive
galaxies are at $z=2.5$. In a hierarchical growth scenario, one expects that the most massive galaxies today have always been the most massive galaxies.  Instead, the expansion model implies that the most massive galaxies at $z \sim 2.5$ evolve into a small fraction of average-mass elliptical galaxies today.  Furthermore, the most massive galaxies in the local universe, with masses $M \gtrsim 3\times10^{11}M_{\odot}$ must then have formed rapidly in the later universe, implying an extremely active merging history of smaller objects.
One might conclude that
they formed through star formation
at lower redshift, but this would be inconsistent
with the stellar ages of massive ellipticals
(e.g., {Thomas} {et~al.} 2005; {van Dokkum} 2008).  

There are
other potential problems
with the physical model proposed by {Fan} {et~al.} (2008).
The growth relies on strong heating of the inner
regions of the galaxy, such as that produced by a central active
galactic nucleus (AGN).  However, the high redshift galaxies in our
sample are already shown to be quiescent, with old stellar
populations.  If there was an active central engine at one point in
the galaxies' histories, it would have already blown out gas and led to expansion of the
galaxy.  While growth through mass loss may have played a
role in the evolution of such galaxies, it is unlikely to do so again
between $z=2.5$ and $z=0$, except possibly through stellar winds and supernovae. Based on simulations of open
clusters, {Fan} {et~al.} (2008) argue that there could be
a long delay between the expulsion of gas and the response of the stellar
distribution to the new potential, but it is not clear whether 
these simulations can easily be applied to massive galaxies.
Finally, the expansion model requires significant fine-tuning of the amount of
mass that is removed from the galaxies: removing a small fraction of the mass does not have an appreciable effect, and removing too much would destroy the galaxies.

Minor mergers (or rather, "un-equal mass mergers") are expected in galaxy formation models, and are
predicted to dominate the mass growth
of massive galaxies at late times (e.g., {Naab} {et~al.} 2007; {Guo} \& {White} 2008).
Simulations have shown that 
the central regions of a galaxy can be minimally affected by
dry mergers but that an  envelope of newly accreted material is
formed that
grows with time ({Naab} {et~al.} 2007).
They have also been observed (e.g., {Schweizer} \& {Seitzer} 1992).  {van Dokkum} (2005) infers that visible
tidal features around nearby elliptical galaxies are caused by
red mergers with median mass ratio $1:4$. It is an open question
whether the merger rate is sufficiently high to produce a factor
of 2--3 growth in mass since $z=2.5$. Models do predict high accretion rates (e.g., {De Lucia} {et~al.} 2006; {Naab} {et~al.} 2007; {Guo} \& {White} 2008), but some observations suggest that mass growth may be small for the highest masses (e.g., {Cool} {et~al.} 2008).

Minor merger models are also qualitatively consistent with the
uniform and gradual evolution of the size -- mass relation
(e.g., {Franx} {et~al.} 2008; {van der Wel} {et~al.} 2008), and
the apparent lack of old massive compact galaxies in the local
Universe (e.g., {Trujillo} {et~al.} 2009). If equal-mass mergers were
a dominant mechanism one might expect 
to find some galaxies that did not experience a major
merger and are therefore left intact at the present day, but
this is very unlikely in a minor merger model.

\subsection{Implied Velocity Dispersions}

\begin{figure}[t]
  \centering
  \includegraphics[scale=0.9]{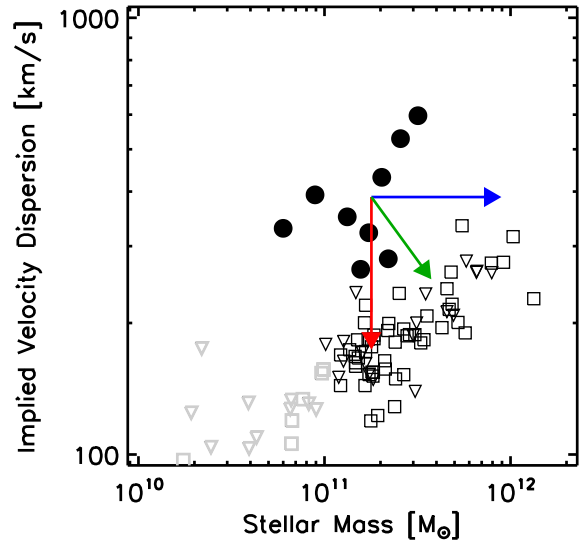}
  \caption{Implied velocity dispersions of high and low redshift galaxies,
along with approximate predictions from the three models discussed
in \S\,4.
}
  \label{fig:dynprop}
\end{figure}

As has been pointed out in several studies, the small sizes and
high masses of the compact red galaxies imply very high velocity
dispersions ({Toft} {et~al.} 2007; {Cimatti} {et~al.} 2008; {van Dokkum} {et~al.} 2008). 
Figure \ref{fig:dynprop} demonstrates the implied dynamical
properties of the nearby and high redshift galaxy population as well
as the possible evolutionary tracks of these galaxies.

We calculated the velocity dispersions from the equation given in
{van Dokkum} \& {Stanford} (2003):
\begin{equation}
\log M \equiv 2 \log \sigma + \log r_e +6.07,
\label{eq:sigma}
\end{equation} 
with $r_e$ in kpc and $M$ in Solar masses. This expression is not very
accurate as it does not take the relation between $M$ and $M/L$ or the
effects of a dark halo into
account, but it does allow a comparison in a self-consistent way.  

Predictions from the simple models of \S\,4 are shown by arrows.
The expansion model predicts that the dispersions decrease over time,
as  the total mass of the galaxy 
remains constant and the effective radius increases.
As discussed in \S\,4.1 the
equal-mass merger model predicts that the 
velocity dispersions remain constant as the mass grows, which
implies that the
descendants have velocity dispersions that are higher than are
implied by the galaxies in the local sample.  Growth by minor
mergers presents a possible method of decreasing the velocity
dispersion of the galaxies, as the expansion is a stronger factor than
mass growth.  This mechanism, again shown by the green arrow on Figure
\ref{fig:dynprop}(a), evolves the compact galaxies onto the velocity
dispersion trend for local galaxies.  This assumes mass growth by a factor of 2.1, based on the same specific minor merging history described in \S\,4.2, and size growth by a factor of 5.  

\section{Summary and Conclusions}

The main result from our study is that nearby elliptical galaxies
have similar average densities within 1\,kpc
as the recently discovered compact ``red and dead'' galaxies at
high redshift.  The descendants of the compact "red and dead" galaxies at z>2 could therefore simply constitute the central parts of today's massive elliptical galaxies.
 
Models dominated by minor mergers (where ``minor''
implies ``not equal mass'') can increase the sizes of the
galaxies efficiently, without violating constraints from the
evolution of the evolution of the
mass function as measured by Marchesini et al.\ (2008).
Interestingly, the evolution
of the mass -- size relation and the mass function together
imply that $\sim 50$\%\footnote{In a minor merger model the exact fraction could range from $\sim 10\% -- 100\%$, depending on the order and mass ratio of mergers.} of elliptical galaxies with mass $\gtrsim 2-3 \times
10^{11}$\,\msun\ may have the remnant of a compact $z=2.5$
galaxy with mass $\gtrsim 1\times 10^{11}$\,\msun\ in its
center.  Models which require energy input by a central engine to ''puff up''
the galaxies can also adequately evolve compact galaxies into
sufficiently common local counterparts, but these models require
significant fine-tuning and may not be physically plausible as the primary growth mechanism.  We note that we did not consider star formation as a way
to grow the compact $z\sim 2.3$
galaxies. Although it is possible that star
formation re-starts at lower redshifts, newly formed stars can only
account for a small fraction of the final mass given the stellar ages
inferred for
massive ($\gtrsim 2 \times 10^11 M_{\odot}$) galaxies at $z=0$
(e.g., Thomas et al 2005, van Dokkum \& van der Marel 2007).
Nevertheless, a small amount of star formation could help
increase the sizes between $z\sim 2.3$ and $z=0$ (see also Franx
et al.\ 2008).

The minor merger model predicts evolution in the
{Magorrian} {et~al.} (1998) relation between
black hole mass and velocity dispersion. The
central black hole will grow
from $z=2.3$ to the present, with the amount
of growth determined by the black hole masses of the infalling galaxies.
However, the velocity dispersion will decrease by a factor of
$\sim 1.5$. Therefore, even if the black hole growth is insignificant,
black hole masses at fixed velocity dispersion
will be significantly lower at $z\sim 2.5$ than at $z=0$.
{Robertson} {et~al.} (2006b) came to the same conclusion
using merger simulations, but this prediction contrasts with
several other studies (e.g., {Cen} 2007; {Woo} {et~al.} 2008).

The galaxy growth models that we describe here are simple and the
empirical findings are preliminary, highlighting the need for further
study. On the modeling side,
the main uncertainties
are whether the merger rate is sufficiently high
to produce the required growth, and whether a realistic
treatment of the dark matter and orbital configurations
retains the high efficiency of minor mergers to ``puff up''
a galaxy. Whatever the dominant physical mechanism turns out
to be, we find that $\alpha \gtrsim 2$ if
the relation between size growth and mass growth is
parameterized as $r_{1+2}/r_1 = (M_{1+2}/M_1)^{\alpha}$.

Inside-out formation via  mergers predicts that stars in the 
central regions of a nearby elliptical galaxy are qualitatively
different from stars at larger radii. Elliptical
galaxies do have color- and metallicity gradients, which could reflect differences in stellar populations between stars formed in-situ and those accreted from other systems (e.g., {Peletier} {et~al.} 1990).  While it is not yet clear
whether these gradients are consistent with such accretion scenarios, it may be difficult to reconcile them with an expansion model alone (see, e.g., {Pipino} \& {Matteucci} 2008).
It is tempting to identify kinematically decoupled cores
(e.g., {Franx} \& {Illingworth} 1988; {Bender} 1988) with
the descendants of the compact galaxies, but the scales of these
features are typically a few 100 pc rather than $\sim 1$\,kpc.
More information on color gradients and
the inner $\sim 1$\,kpc of the compact
high redshift galaxies will provide important additional constraints.

Our determinations of stellar density profiles and masses can  be
improved. The calculated density profiles and integrated masses are based on Sersic
profile fits to the galaxy light distributions, not on the actual
light profiles themselves.  Furthermore, for the high
redshift galaxies the profiles within $\sim 1$\,kpc are extrapolations,
as the galaxies are not resolved on smaller scales.
The conversion from light to mass is also very uncertain.
The conversion for the local samples
ignores scatter in the $M/L$ versus $L$ relation, and ignores
gradients in $M/L$ ratio.
The mass estimates of the high redshift galaxies
are based on stellar population models and are very sensitive to
the assumed IMF and to possible contributions from dark matter.
As noted in \S\,1,
bottom-light IMFs would change the masses and alter the
required amount of size- and mass evolution to bring the galaxies
to local relations. 

Measurements of absorption-line kinematics of high redshift
compact galaxies would provide a direct test of the IMF, and of
several of the other assumptions that enter the analysis
(see, e.g., {Cimatti} {et~al.} 2008). van der Wel
et al.\ (2008) find that the observed
size evolution at $0<z<1$ is similar when
dynamical masses rather than photometric masses are used, but this
needs to be verified at higher redshifts.

\begin{acknowledgements}
We thank Avishai Dekel and Jeremiah Ostriker for helpful discussions.  Support from NASA grants HST-GO-10808 and HST-GO-10809 is gratefully acknowledged.
\end{acknowledgements}

%\bibliography{}

%% --------------------------------------------------------------------
%% Fri Jan 30 15:32:56 2009
%%   This file was generated automagically from the files
%%   submit.bbl and submit.tex using
%%     nat2jour.pl
%%   This file should accompany submit-aas.tex.
%% --------------------------------------------------------------------

\end{document}